\documentclass[twocolumn]{autart}

\pdfoutput=1

\usepackage{graphicx}          

\usepackage{acronym}
\usepackage{color}
\usepackage{amsmath} 
\usepackage{amssymb} 
\usepackage{subfigure}
\usepackage{enumitem}  
\usepackage{algorithmic}
\usepackage{psfrag}  
\usepackage[process=auto]{pstool}
\usepackage{multirow} 
\usepackage{url} 
\usepackage{accents} 

\begin{document}

\acrodef{FDIR}{Fault Detection, Isolation and Reconstruction}
\acrodef{FDIE}{Fault Detection, Isolation and Estimation}
\acrodef{FDD}{Fault Detection and Diagnosis}
\acrodef{FTC}{Fault Tolerant Control}
\acrodef{FDI}{Fault Detection and Isolation}
\acrodef{IMU}{Inertial Measurement Unit}
\acrodef{ADS}{Air Data Sensors}
\acrodef{GPS}{Global Positioning Systems}
\acrodef{SMO}{Sliding Mode Observers}
\acrodef{DOS}{Dedicated Observer Scheme}
\acrodef{KF}{Kalman Filter}
\acrodef{EKF}{Extended Kalman Filter}
\acrodef{IEKF}{Iterated Extended Kalman Filter}
\acrodef{UKF}{Unscented Kalman Filter}
\acrodef{AUKF}{Augmented Unscented Kalman Filter}
\acrodef{MM}{multiple model}
\acrodef{DMAE}{Double-Model Adaptive Estimation}
\acrodef{DMAE-NSR}{Double-Model Adaptive Estimation-No Selective Reinitialization}
\acrodef{MMAE}{Multiple-Model Adaptive Estimation}
\acrodef{IMM}{Interacting Multiple-Model}
\acrodef{AMMAE}{Augmented Multiple-Model Adaptive Estimation}
\acrodef{SRMMAE}{Selective-Reinitialisation Multiple Model Adaptive Estimation}
\acrodef{FF}{Fault Free}
\acrodef{SR}{Selective-Reinitialization}
\acrodef{ADDSAFE}{Advanced Fault Diagnosis for Sustainable Flight Guidance and Control}

\acrodef{RTSKF}{Robust Three-Step Kalman Filter}
\acrodef{RMSE}{root mean square error}
\acrodef{OTSKF}{Optimal Two-Stage Kalman Filter}

\begin{frontmatter}

\title{Framework for state and unknown input estimation of linear time-varying systems\thanksref{footnoteinfo}}

\thanks[footnoteinfo]{This paper was not presented at any IFAC 
meeting. Corresponding author Peng Lu. Tel. +31 152783466. 
Fax +31 152786480.}

\author[cs]{Peng Lu}\ead{P.Lu-1@tudelft.nl},    
\author[cs]{Erik-Jan van Kampen}\ead{E.vanKampen@tudelft.nl},               
\author[cs]{Cornelis C. de Visser}\ead{c.c.devisser@tudelft.nl},   
\author[cs]{Qiping Chu}\ead{q.p.chu@tudelft.nl}  

\address[cs]{Delft University of Technology, Kluyverweg 1, 2629HS Delft, The Netherlands}

\begin{keyword}                           
Kalman filtering; state estimation; unknown input filtering; fault estimation; Double-Model Adaptive Estimation.               
\end{keyword}                             

\begin{abstract}                          
The design of unknown-input decoupled observers and filters requires the assumption of an existence condition in the literature. This paper addresses an unknown input filtering problem where the existence condition is not satisfied. Instead of designing a traditional unknown input decoupled filter, a Double-Model Adaptive Estimation approach is extended to solve the unknown input filtering problem. It is proved that the state and the unknown inputs can be estimated and decoupled using the extended Double-Model Adaptive Estimation approach without satisfying the existence condition.
Numerical examples are presented in which the performance of the proposed approach is compared to methods from literature.
\end{abstract}

\end{frontmatter}

\section{Introduction}
\label{s:1}
Faults and model uncertainties such as disturbances can be represented as unknown inputs. The problem of filtering in the presence of unknown inputs has received intensive attention in the past three decades. 

It is common to treat the unknown inputs as part of the system state and then estimate the unknown inputs as well as the system state \cite{Heish1999}. This is an augmented Kalman filter, whose computational load may become excessive when the number of the unknown inputs is comparable to the states of the original system \cite{Friedland1969}. 
Friedland \cite{Friedland1969} derived a two-stage Kalman filter which decomposes the augmented filter into two reduced-order filters. However, Friedland's approach is only optimal in the presence of a constant bias \cite{Heish1999}. Hsieh and Chen derived an optimal two-stage Kalman filter which performance is also optimal for the case of a random bias \cite{Heish1999}.

On the other hand, unknown input filtering can be achieved by making use of unbiased minimum-variance estimation \cite{Hsieh2009,Kitandis1987,Darouach1997,Hou1998,Hsieh2000,Chen1996}. 
Kitanidis \cite{Kitandis1987} first developed an unbiased recursive filter based on the assumption that no prior information about the unknown input is available \cite{Gillijns2007}. Hou and Patton \cite{Hou1998} used an unknown-input decoupling technique and the innovation filtering technique to derive a general form of unknown-input decoupled filters \cite{Hou1998,Hsieh2000}. 
Darouach, Zasadzinski and Boutayeb \cite{Darouach2003} extended Kitanidis' method using a parameterizing technique to derive an optimal estimator filter. The problem of joint input and state estimation, when the unknown inputs only appear in the system equation, was addressed by Hsieh \cite{Hsieh2000} and Gillijns and De Moor \cite{Gillijns2007c}. Gillijns and De Moor \cite{Gillijns2007} further proposed a recursive three-step filter for the case when the unknown inputs also appear in the measurement equation. However, their approach requires the assumption that the distribution matrix of the unknown inputs in the measurement equation is of full rank. Cheng et al. \cite{Cheng2009} proposed a global optimal filter which removed this assumption, but this filter is limited to state estimation \cite{BenHmida2010}. Later, Hsieh \cite{Hsieh2010a} presented a unified approach to design a specific globally optimal state estimator which is based on the desired form of the distribution matrix of the unknown input in the measurement equation \cite{Hsieh2010a}.

However, all the above-mentioned filters require the assumption that an existence condition is satisfied. This necessary condition is given by Hou and Patton \cite{Hou1998} and Darouach, Zasadzinski and Boutayeb \cite{Darouach2003}, in the form of rank condition (\ref{e:existence condition}). Hsieh \cite{Hsieh2010a} presents different decoupling approaches for different special cases. However, these approaches also have to satisfy the existence condition \eqref{e:existence condition}.
In some applications, such as that presented in the current paper, the existence condition is not satisfied. Therefore, a traditional unknown input decoupled filter can not be designed. 

Recently, particle filters are also applied to unknown input estimation \cite{Gordan1993,Doucet2000,Verma2004}. These filters can cope with systems with non-Gaussian noise and have a number of applications such as for robot fault detection \cite{Caron2007,Freitas2002,Zhao2014a}. In this paper, the performance of unknown input estimation using particle filters will be compared with that of our approach.

This paper proposes an extended \ac{DMAE} approach, which can cope with the unknown input filtering problem when a traditional unknown input filter can not be designed.  
The original \ac{DMAE} approach, which was proposed by Lu et al. \cite{Lu2015} for the estimation of unknown inputs in the measurement equation, is extended to allow estimation of the unknown inputs which appear both in the system equation and the measurement equation. The unknown inputs are augmented as system states and are modeled as random walk processes. The unknown inputs in the system equation are assumed to be Gaussian random processes of which covariances are estimated on-line. It is proved that the state and unknown inputs can be estimated and decoupled while not requiring the existence condition. Two illustrative examples are given to demonstrate the effectiveness of the proposed approach with comparison to other methods from literature such as the \ac{RTSKF} \cite{Gillijns2007}, the \ac{OTSKF} \cite{Heish1999} and the particle filters \cite{Gordan1993,Doucet2000}.

The structure of the paper is as follows: the preliminaries of the paper are given in Section~\ref{s:2}, formulating the filtering problem when the existence condition for a traditional unknown input decoupled filter is not satisfied and generalizing the \ac{DMAE} approach. In Section~\ref{s:3}, the extension of the \ac{DMAE} approach to the filtering problem when the unknown inputs appear both in the system equation and the measurement equation is presented. Furthermore, the on-line estimation of the covariance matrix of the unknown inputs is introduced. It is proved that the state and the unknown inputs can still be estimated and decoupled in Section~\ref{s:4}. In Section~\ref{s:5}, two illustrative examples are given to show the performance of the proposed approach with comparison to some existing unknown-input decoupled filters. Finally, Section~\ref{s:6} concludes the paper.

\section{The \ac{DMAE} approach}
\label{s:2}
This section presents the problem formulation and the \ac{DMAE} approach.

\subsection{Problem formulation}
\label{s:21}
Consider the following linear time-varying system:
\begin{align}
\label{e:xdot}
x_{k+1} &= A_k x_k + B_k u_k + E_k d_k + w_k\\
\label{e:y}
y_{k} &= H_k x_k + F_k f_k + v_k
\end{align}
where $x_k \in \mathbb{R}^n$ represents the system states, $y_k\in \mathbb{R}^m$ the measurements, $d_k$ and $f_k$ are the unknown inputs. Specifically, $d_k\in \mathbb{R}^{n_d}$ the disturbances, $f_k\in \mathbb{R}^{n_f}$ are the output faults. $w_k$ and $v_k$ are assumed to be uncorrelated zero-mean white noise sequences with covariance $Q_k$ and $R_k$ respectively. $u_k$, the known inputs, is omitted in the following discussion because it does not affect the filter design \cite{Hou1998}. Without loss of generality, we consider the case: $n=m=n_d=n_f$ and $\text{rank}\ H_k=\text{rank}\ E_k=\text{rank}\ F_k = m$, which implies all the states are influenced by $d_k$ and $f_k$. It should be noted that the approach proposed in this paper can be readily extended to the case when $n\neq m$ or $\text{rank}\ H_k \neq \text{rank}\ E_k$. 

The unknown inputs are denoted as $d_k'$, i.e., $d_k' = \begin{bmatrix}
d_k\\f_k
\end{bmatrix} \in \mathbb{R}^{n_{d'}}$. Then, model~(\ref{e:xdot}) and (\ref{e:y}) can be reformulated into the general form as given in Hou and Patton \cite{Hou1998} and Darouach, Zasadzinski and Boutayeb \cite{Darouach2003}:
\begin{align}
\label{e:xdot 2}
x_{k+1} &= A_k x_k + E'_k d'_k + w_k\\
\label{e:y 2}
y_{k} &= H_k x_k + F'_k d'_k + v_k
\end{align}
In this paper, $E'_k = [E_k\ 0]$, $F'_k = [0\ F_k]$. The existence of an unknown-input decoupled filter must satisfy the following existence condition \cite{Hou1998,Darouach2003}:

\begin{equation}
\label{e:existence condition}
\text{rank} 
\begin{bmatrix}
F'_k& H_kE'_k\\
0& F'_k
\end{bmatrix}
=
\text{rank}\ [F'_k]  +
\text{rank}
\begin{bmatrix}
E'_k\\F'_k
\end{bmatrix} 
\end{equation}

In our case, since $\text{rank}\ H_k = m$, the left-hand side of condition~(\ref{e:existence condition}) is $2m$ while the right-hand side is $3m$. Therefore, the above existence condition does not hold, which means that all the unknown-input filters mentioned in the introduction can not be directly implemented.

In this paper, we consider the consecutive bias fault estimation of a system subjected to disturbances, as described in Eqs.~\eqref{e:xdot} and \eqref{e:y}. Although the existence condition of designing a traditional unknown input decoupled filter is not satisfied, it will be shown that the unknown inputs can still be decoupled using an extended \ac{DMAE} approach.

\textbf{Remark 1}. The model described by Eqs.~\eqref{e:xdot} and \eqref{e:y} is useful for applications where the disturbances appear in the system equation and the faults appear in the measurement equation, such as bias fault estimation in aircraft air data sensors \cite{Lu2015}.

\subsection{The \ac{DMAE} approach}
\label{s:22}
The \ac{DMAE}1 approach proposed in Lu et al. \cite{Lu2015} considers the model~(\ref{e:xdot}) and ~(\ref{e:y}) for $d_k=0$ ($n_d=0$). It is referred to as the \ac{DMAE} approach in this paper, which is generalized in the following.

The \ac{DMAE} \cite{Lu2015}, which is a modified approach of multiple-model-based approach \cite{Magill1965,Maybeck1999}, is composed of two \acp{KF} operating in parallel: a no-fault (or fault-free) filter and an augmented fault filter. These two filters are based on two modes of the system: fault-free ($f_k=0$) and faulty ($f_k\neq 0$). The two filters use the same vector of measurements $Y$ and vector of input $u$, and are based on the same equations of motion, while each hypothesizes a different fault scenario. The state vector of the no-fault filter $x_{n\!f}$ and that of the augmented fault filter $x_{a\!f}$ are as follows:
\begin{align}
x_{n\!f,k} &= x_k ,\
x_{a\!f,k} = \begin{bmatrix}
x_{n\!f,k}\\ f_k
\end{bmatrix}
\end{align}
where ``$n\!f$'' means no fault and ``$a\!f$'' means augmented fault. It can be noted that the state vector of the augmented fault filter is the state vector of the no-fault filter with augmentation of the fault vector $f_k$.

\begin{figure}
\includegraphics[width=0.5\textwidth]{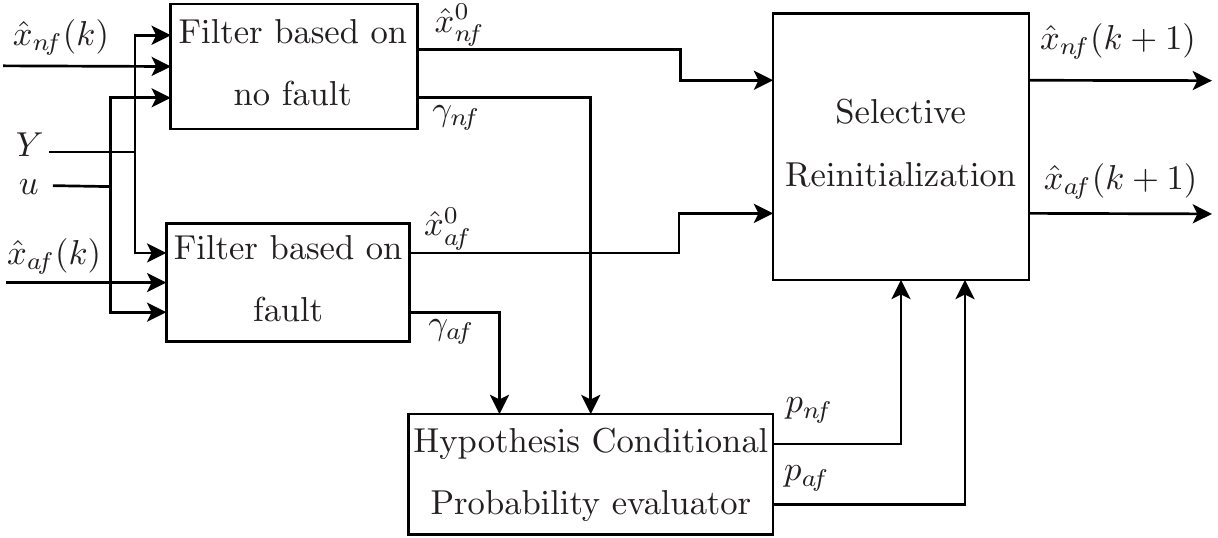}{}
\caption{Block diagram for the DMAE approach}
\label{f:DMAE block}
\end{figure}

At time step $k$, each of the filters produces a state estimate $\hat{x}_i^0(k)$ and a vector of innovations $\gamma_i(k)$. The principle is that the \ac{KF} which produces the most well-behaved innovations, contains the model which matches the true faulty model best \cite{Magill1965,Maybeck1999}. The block diagram of the \ac{DMAE} is given in Fig.~\ref{f:DMAE block}. 

\begin{figure}
\includegraphics[width=0.5\textwidth]{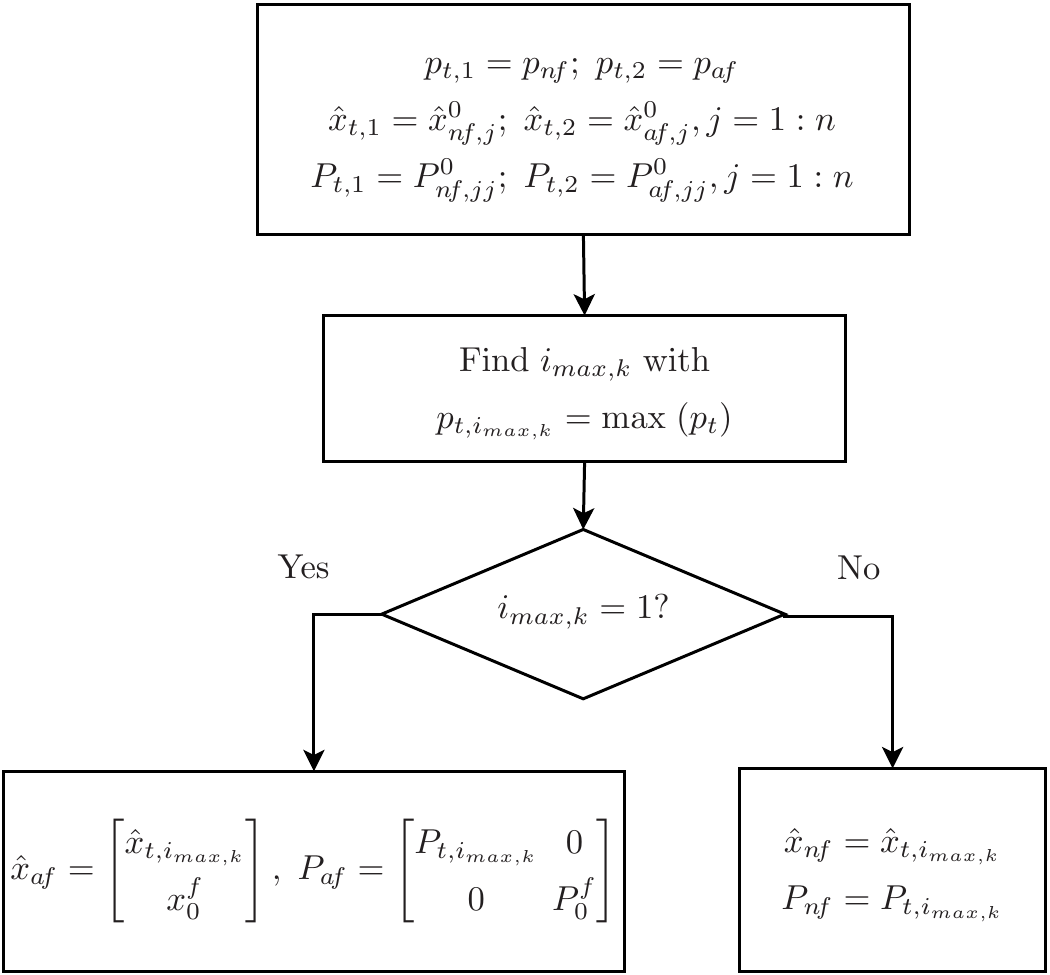}{}
\caption{Flow chart of the Selective Reinitialization algorithm. Note $n$ refers to the dimension of $\hat{x}_{n\!f}$.}
\label{f:SR chart}
\end{figure}

A hypothesis test uses the innovation $\gamma_i(k)$ and the innovation covariance matrix $C_i(k)$ of the filters in order to assign a conditional probability to each of the filters. Let $a$ denote the fault scenarios of the system. If we define the hypothesis conditional probability $p_i(k)$ as the probability that $a$ is assigned $a_i$ for $i=1,2$ ($a_1=n\!f$, $a_2=a\!f$), conditioned on the measurement history up to time step $k$:
\begin{equation}
p_i(k)=\text{Pr}[a=a_i|Y(k)=Y_k],\ \ \ i=1,2
\end{equation}
then the conditional probability of the two filters can be updated recursively using the following equation:
\begin{equation}
\label{e:pdf compute}
p_i(k)=\frac{f_{y_k|a,Y_{k-1}}(y_k|a_i,Y_{k-1})p_i(k-1)}{\sum\limits_{j=1}^2f_{y_k|a,Y_{k-1}}(y_k|a_j,Y_{k-1})p_j(k-1)},\ \ \ i=1,2
\end{equation}  
where $Y_{k-1}$ is the measurement history vector which is defined as $Y_{k-1}=\{y(1),y(2),..,y(k-1)\}$.\\ $f_{y_k|a,Y_{k-1}}(y_k|a_i,Y_{k-1})$ is the probability density function which is given by the following Gaussian form \cite{Maybeck1999}:
\begin{align}
\label{e:density function}
f_{y(k)|a,Y_{k-1}}&(y(k)|a_i,Y_{k-1}) \nonumber \\ 
= &\beta_i(k)\ \text{exp}\{-\gamma_i^T(k)C^{-1}_i(k)\gamma_i(k)/2\}
\end{align}
where 
\begin{equation}
\label{e:beta i}
\beta_i(k)=\frac{1}{(2\pi)^{m/2}|C_i(k)|^{1/2}}
\end{equation}
In Eq.~(\ref{e:beta i}), $|\bullet|$ denotes the determinant of the covariance matrix $C_i(k)$ which is computed by the \ac{KF} at time step $k$. The filter which matches the fault scenario produces the smallest innovation which is the difference between the estimated measurement and the true measurement. Therefore, the conditional probability of the filter which matches the true fault scenario is the highest between the two filters. After the computation of the conditional probability, the state estimate of the nonlinear system $\hat{x}(k)$ can be generated by the weighted state estimate $\hat{x}_i(k)$ of the two filters: 
\begin{align}
\hat{x}(k)&= \sum\limits_{i=1}^2\hat{x}_i(k)p_i(k) \nonumber \\
&= \hat{x}_{n\!f}(k)p_{n\!f}(k) + \hat{x}_{a\!f}(k)p_{a\!f}(k).
\label{e:state pdf}
\end{align}
The fault is only estimated by the augmented fault filter and the estimate is denoted as $\hat{f}(k)$. The probability-weighted fault estimate of the \ac{DMAE} approach $\bar{f}(k)$ is calculated as follows:
\begin{equation}
\label{e:fault pdf}
\bar{f}(k)=\hat{f}(k)p_{a\!f}(k)
\end{equation}
The core of the \ac{DMAE} approach is selective reinitialization. The flow chart of the selective reinitialization algorithm is presented in Fig.~\ref{f:SR chart}.

In the algorithm, $\hat{x}_{n\!f}^0$ ($\hat{x}_{a\!f}^0$) and $\hat{x}_{n\!f}$ ($\hat{x}_{a\!f}$)  denote the state estimate of the no-fault (augmented fault) filter before and after the reinitialization, respectively. $P_{n\!f}^0$ ($P_{a\!f}^0$) and $P_{n\!f}$ ($P_{a\!f}$)  denote the covariance of state estimate error of the no-fault (augmented fault) filter before and after the reinitialization, respectively. $\hat{x}_t$, $p_t$ and $P_t$ are the vectors which contain the state estimate, model probability and the covariance matrix of state estimation error of the no-fault filter and the fault filter respectively. $i_{max,k}$ is the index of the model with the maximum model probability at time step $k$. ${x}_0^f$ and $P_0^f$ are the parameters which are used for the initialization of the fault filter.

\section{Extension of the \ac{DMAE} approach}
\label{s:3}
The \ac{DMAE} approach can achieve an unbiased estimation of $x_k$ and $f_k$ when $d_k=0$ \cite{Lu2015}. However, when $d_k\neq 0$, the unknown-input filtering problem becomes more challenging. Since the existence condition~(\ref{e:existence condition}) is no longer satisfied, traditional unknown-input decoupled filters can not be designed.

In this section, the \ac{DMAE} is extended to the case when $d_k\neq 0$. In order to achieve this, the state vectors of the no-fault filter and augmented fault filter are changed to:
\begin{align}
\bar{x}_{n\!f,k} = \begin{bmatrix}
x_k\\ d_k
\end{bmatrix},\ 
\bar{x}_{a\!f,k} = \begin{bmatrix}
\bar{x}_{n\!f,k} \\ f_k
\end{bmatrix} 
\end{align}
where $\bar{x}_{n\!f,k} \in \mathbb{R}^{n+n_d}$ and $\bar{x}_{a\!f,k} \in \mathbb{R}^{n+n_d+n_f}$.
The state vector of the augmented fault filter is that of the no-fault filter augmented with the fault vector. Therefore, the state vector of the no-fault filter can be inferred from that of the augmented fault filter and vice versa. 

The random walk process provides a useful and general tool for the modeling of unknown time-varying processes \cite{Friedland1969,Park2000,Hsieh2000}.
$d_k$ can be modeled by a random walk process \cite{Park2000,Hsieh2000} as:
\begin{align}
d_{k+1} = d_{k}+w_{d,k},
\end{align}  
where $w_{d,k}$ is a white noise sequence with covariance: $E\{w_{\!d,k}(w_{\!d,l})^T\} = Q_k^{d} \delta_{kl}$. $f_k$ is also modeled as a random walk process as:
\begin{align}
f_{k+1} = f_{k}+w_{f,k}, 
\end{align} 
where $w_{f,k}$ is a white noise sequence with covariance: $E\{w_{\!f,k}(w_{\!f,l})^T\} = Q_k^{f} \delta_{kl}$.
Then, the system model and measurement model of the no-fault filter can be described as follows:
\begin{align}
\label{e:xdot no fault}
\bar{x}_{n\!f,k+1} &= \bar{A}_{n\!f,k} \bar{x}_{n\!f,k} + \bar{w}_{n\!f,k}\\
\label{e:y no fault}
y_k &= \bar{H}_{n\!f,k} \bar{x}_{n\!f,k} + v_k
\end{align}
where
\begin{align}
\bar{A}_{n\!f,k} = \begin{bmatrix}
A_k & E_k\\
0 & I
\end{bmatrix},  
\bar{H}_{n\!f,k} = [H_k\ 0], 
\label{e:wb no fault}
\bar{w}_{n\!f,k} = \begin{bmatrix}
w_k \\ w_{\!d,k}
\end{bmatrix}
\end{align}
The model of the augmented fault filter is as follows:
\begin{align}
\label{e:xdot fault}
\bar{x}_{a\!f,k+1} &= \bar{A}_{a\!f,k} \bar{x}_{a\!f,k} + \bar{w}_{a\!f,k}\\
\label{e:y fault}
y_k &= \bar{H}_{a\!f,k} \bar{x}_{a\!f,k} + v_k
\end{align}
where
\begin{align}
\bar{A}_{a\!f,k} = \begin{bmatrix}
\bar{A}_{n\!f,k} & 0\\
0 & I
\end{bmatrix},  
\bar{H}_{a\!f,k} = [\bar{H}_{n\!f,k} \ F_k] ,
\label{e:wb fault}
\bar{w}_{a\!f,k} = \begin{bmatrix}
\bar{w}_{n\!f,k} \\ w_{\!f,k}
\end{bmatrix}
\end{align}
Since the difference from the \ac{DMAE} in Lu et al. \cite{Lu2015} is the augmentation of $d_k$, only the covariance related to $w_{\!d,k}$, i.e., $Q_k^{d}$ is discussed below. 
It should be noted that $Q_k^{d}$ is usually unknown, the optimality of the filter can be compromised by a poor choice of $Q_k^{d}$ \cite{Kitandis1987,Hsieh2000}. If $Q_k^{d}$ is not properly chosen, it can influence the estimation of $d_k$ as well as $x_k$.

This paper proposes a method to adapt $Q_k^{d}$ by making use of the augmented fault filter of the \ac{DMAE} approach. 
To compensate for the effect of a bad choice of $Q_k^{d}$ on the estimation of $x_k$,
the system noise vector $\bar{w}_{n\!f,k}$ in Eqs.\eqref{e:xdot no fault}, \eqref{e:wb no fault} and \eqref{e:wb fault} is modified to:
\begin{align}
\bar{w}_{n\!f,k} = \begin{bmatrix}
w_k + w'_k \\ w_{d,k}
\end{bmatrix}
\end{align}
where $w'_k$ is the noise used to compensate for the effect of a bad choice of $Q_k^{d}$ on the estimation of $x_k$. In this paper, we approximate $w'_k$ by $E_k w_{d,k}$. Therefore, $\bar{w}_{n\!f,k}$ is
\begin{align}
\bar{w}_{n\!f,k} = \begin{bmatrix}
w_k + E_k w_{d,k} \\ w_{d,k}
\end{bmatrix}
\end{align}

Let $\hat{\bar{x}}_{a\!f,k-1|k-1}$ denote the unbiased estimate of $\bar{x}_{a\!f,k-1}$ given measurements up to time $k-1$. $\hat{x}_{k-1|k-1}$, $\hat{d}_{k-1|k-1}$ and $\hat{f}_{k-1|k-1}$ denote the estimates of $x_{k-1}$, $d_{k-1}$ and $f_{k-1}$, respectively. The innovation of the augmented fault filter is:
\begin{align}
\gamma_{a\!f,k} & = y_k - \bar{H}_{a\!f,k} \hat{\bar{x}}_{a\!f,k|k-1} \nonumber \\
& = H_k A_{k-1}\tilde{x}_{k-1|k-1} + H_k E_{k-1} \tilde{d}_{k-1|k-1} + F_k \tilde{f}_{k-1|k-1} \nonumber \\
&   + H_k w_{k-1} + H_k E_{k-1}w_{d,k-1}  + F_k w_{f,k-1} + v_k
\end{align}
with
\begin{align}
\tilde{x}_{k-1|k-1} &:= x_{k-1} - \hat{x}_{k-1|k-1}\\
 \tilde{d}_{k-1|k-1} &:= d_{k-1} - \hat{d}_{k-1|k-1}\\
 \tilde{f}_{k-1|k-1} &:= f_{k-1} - \hat{f}_{k-1|k-1}
\end{align}
Therefore, the innovation covariance of the augmented fault filter is:
\begin{align}
C_{a\!f,k} &= E\{ \gamma_{a\!f,k} \gamma_{a\!f,k}^T \}  \nonumber \\
&= H_k A_{k-1}P^x_{k-1|k-1}A_{k-1}^T H_k^T  \nonumber \\
&+ H_k E_{k-1}P^d_{k-1|k-1} E^T_{k-1} H_k^T + F_k P^f_{k-1|k-1} F^T_k \nonumber \\
&+ H_k A_{k-1}P^{xd}_{k-1|k-1}E_{k-1}^TH_k^T + H_k A_{k-1}P^{xf}_{k-1|k-1}F_{k}^T \nonumber \\
&+ H_k E_{k-1}P^{dx}_{k-1|k-1}A_{k-1}^T H_k^T + H_k E_{k-1}P^{df}_{k-1|k-1}F_{k-1}^T \nonumber \\
& + F_{k-1}P^{fx}_{k-1|k-1}A_{k-1}^T H_k^T + F_k P^{fd}_{k-1|k-1}E_{k-1}^T H_k^T +  R_k \nonumber \\
&  + H_k Q_{k-1} H_k^T + H_k E_{k-1} Q_{k-1}^d E_{k-1}^T H_k^T + F_k Q_{k}^f F_k^T 
\end{align} 
where the covariance matrices are defined as follows:
\begin{align}
P^x_{k|k} &:= E[\tilde{x}_{k|k} \tilde{x}^T_{k|k}  ] ,\
P^d_{k|k} := E[\tilde{d}_{k|k} \tilde{d}^T_{k|k}  ] \nonumber \\
P^f_{k|k} &:= E[\tilde{f}_{k|k} \tilde{f}^T_{k|k} ] ,\
P^{xd}_{k|k} := E[\tilde{x}_{k|k} \tilde{d}^T_{k|k}  ] \nonumber \\
P^{dx}_{k|k} &:= E[\tilde{d}_{k|k} \tilde{x}^T_{k|k}  ] ,\
P^{xf}_{k|k} := E[\tilde{x}_{k|k} \tilde{f}^T_{k|k}  ] \nonumber \\
P^{fx}_{k|k} &:= E[\tilde{f}_{k|k} \tilde{x}^T_{k|k}  ] ,\
P^{df}_{k|k} := E[\tilde{d}_{k|k} \tilde{f}^T_{k|k}  ] \nonumber \\
P^{fd}_{k|k} &:= E[\tilde{f}_{k|k} \tilde{d}^T_{k|k}  ] \nonumber . 
\end{align}
The actual $C_{a\!f,k}$ is approximated as follows \cite{Mehra1970,Xia1994}:
\begin{align}
\label{e:innovation covariance}
\hat{C}_{a\!f,k} = \frac{1}{N} \sum\limits^{k}_{j=k-N+1} \gamma_{a\!f,j} \gamma_{a\!f,j}^T
\end{align}
$Q_k^d$ can be approximated by the main diagonal of
\begin{align}
\label{e:Qd compensation}
E_{k-1}^{-1}H_k^{-1} \tilde{Q}_k (H_k^T)^{-1} (E_{k-1}^T)^{-1} 
\end{align}
with $\tilde{Q}_k$ is a diagonal matrix defined as:
\begin{align}
\label{e:Qd non-negative}
\tilde{Q}_k := \text{diag}( \text{max}\{0,\hat{Q}_{k,11}\}, ..., \text{max}\{0,\hat{Q}_{k,mm}\} ) \end{align}
where $\hat{Q}_{k,jj}, j = 1,2,...,m$ is the $j$th diagonal element of $\hat{Q}_k$ which is denoted as:
\begin{align}
\hat{Q}_k &= (\hat{C}_{a\!f,k} - H_k Q_{k-1} H_k^T - F_k Q_{k}^f F_k^T - R_k)
\end{align}
The restriction $\tilde{Q}_{k,jj} \geq 0,j=1,2,...,m$ in Eq.~\eqref{e:Qd non-negative} is to preserve the properties of a variance \cite{Jazwinski1969}.

\section{Unknown input decoupled filtering }
\label{s:4}
This section proves that the unknown input decoupled filtering can be achieved using the extended \ac{DMAE} approach which does not need to satisfy the existence condition (\ref{e:existence condition}). Let $l$ ($l\ge 1$) denote the time step when the first fault occurs and $l_e$ denote the time step when the first fault is removed, which means $f_k=0$ when $k<l$ and $f_k \neq 0$ when $l\leq k\leq l_e$. Without loss of generality, it will be proven that $f_k$ can be estimated when $k\leq l_e$.

\subsection{Unknown input estimation during $k<l$}
\begin{thm}
During $k<l$, an unbiased estimate of $d_k$ can be achieved by the fault-free filter  of the extended \ac{DMAE} approach.
\end{thm}
\begin{pf}
When $k<l$, $f_k=0$. The fault-free model matches the true fault scenario while the augmented fault filter does not. Therefore, according to the \ac{DMAE} approach, $i_{max,k}=1$ during this time period. 

The system model during this period is as follows:
\begin{align}
\label{e:xdot ff}
x_{k+1} &= A_k x_k + E_k d_k + w_k\\
\label{e:y ff}
y_{k} &= H_k x_k + v_k
\end{align}
Under this situation, $d_k$ can be estimated using the fault-free filter whose convergence condition will be discussed later. 
\rlap{$\qquad \Box$}
\end{pf}

The estimation of $d_k$ and $f_k$ when $l\leq k \leq l_e$ will be discussed in the following.

\subsection{Unknown input estimation at $k=l$}

For the sake of readability, the subscript ``$a\!f$'' will be discarded for the remainder of the section. All the variables with a bar on top in the remainder of this section refer to the augmented fault filter.

Using the \ac{DMAE} approach, the Kalman gain $\bar{K}_l$ can be partitioned as follows:
\begin{align}
\bar{K}_l = \begin{bmatrix}
K_l^x\\ K_l^d\\ K_l^f
\end{bmatrix}
\end{align}
where $K_l^x$, $K_l^d$ and $K_l^f$ are the Kalman gains associated with $x_k$, $d_k$ and $f_k$, respectively.

\begin{lem}
\label{l:1}
Let $\hat{x}_{l-1|l-1}$ and $\hat{d}_{l-1|l-1}$ be unbiased, if $x^f_0$ is chosen to be $0$ or sufficiently small, then $f_l$ can be estimated by the augmented fault filter if and only if $K_l^f$ satisfies
\begin{align}
K_l^f F_l = I.
\end{align}
\end{lem}

\begin{pf}
The innovation of the augmented filter is
\begin{align}
\bar{\gamma}_{l} 
\label{e:gamma af}
& = e_l + F_l {f}_{l} 
\end{align}
where $e_l$ is defined as
\begin{align}
\label{e:ei}
e_l &:= H_l A_{l-1}\tilde{x}_{l-1|l-1} + H_l E_{l-1} \tilde{d}_{l-1|l-1}  \nonumber \\
 &+  H_l w_{l-1} + H_l E_{l-1}w_{d,l-1} + v_l
\end{align}
Since $\hat{x}_{l-1|l-1}$ and $\hat{d}_{l-1|l-1}$ are unbiased (this can be achieved by the \ac{DMAE}1 in Lu et. al \cite{Lu2015} since $f_k=0$ when $k< l$), $E[ e_l ] =0$. 
 
Consequently, the expectation of $\bar{\gamma}_l$ is:
\begin{align}
 E[ \bar{\gamma}_l ] = F_l f_l. 
\end{align}

The estimation of the fault can be given by
\begin{align}
\label{e:fault estimation af}
\hat{f}_{l|l} &= \hat{f}_{l|l-1} + K_l^f \bar{\gamma}_l \nonumber \\
&=  \hat{f}_{l-1|l-1} + K_l^f \bar{\gamma}_l 
\end{align}
Since $i_{max,k}=1$ when $k<l$, according to the flow chart of the selective reinitialization algorithm given in Fig.~\ref{f:SR chart}, Eq.~(\ref{e:fault estimation af}) can be further written into
\begin{align}
\label{e:fault estimation af 2}
\hat{f}_{l|l} &= x_0^f + K_l^f \bar{\gamma}_l 
\end{align}

Substituting (\ref{e:gamma af}) into (\ref{e:fault estimation af 2}), yields
\begin{align}
\label{e:fi est}
\hat{f}_{l|l} = K_l^f F_l f_l + K_l^f e_l
\end{align}

Consequently, the expectation of $\hat{f}_{l|l}$
\begin{align}
E[ \hat{f}_{l|l} ] = E[ K_l^f F_l f_l ].
\end{align}

Therefore, it can concluded that $f_l$ can be estimated if and only if $K_l^f$ satisfies 
\begin{align}
\label{e:Kif}
K_l^f F_l = I. \rlap{$\qquad \Box$}
\end{align}

\end{pf}

\begin{thm}
\label{t:1}
Let $\hat{x}_{l-1|l-1}$ and $\hat{d}_{l-1|l-1}$ be unbiased, then $f_l$ can be estimated by the augmented fault filter of the \ac{DMAE} approach by choosing a sufficiently large $P_0^f$ and a sufficiently small $x_0^f$.
\end{thm}
\begin{pf}
Define the following covariance matrix:
\begin{align}
\bar{P}_{l-1|l-1} &:= E[ \tilde{ \bar{x} }_{l-1|l-1} \  \tilde{ \bar{x} }_{l-1|l-1}^T] \nonumber 
\end{align}
where $\tilde{ \bar{x} }_{l-1|l-1} = \bar{x}_{l-1} - \hat{ \bar{x} }_{l-1|l-1}$.

Due to the selective reinitialization algorithm given in Fig.~\ref{f:SR chart}, $P_{l-1|l-1}^f = P_0^f$. Therefore, the covariance of the state prediction error $\bar{P}_{l|l-1}$ can be computed and partitioned as follows:
\begin{align}
\bar{P}_{l|l-1} &= \bar{A}_{l-1}
\begin{bmatrix}
P_{l-1|l-1}^x & P_{l-1|l-1}^{xd} &0\\
P_{l-1|l-1}^{dx} & P_{l-1|l-1}^d &0\\
0 &0 & P_{0}^f
\end{bmatrix}
\bar{A}_{l-1}^T \nonumber \\
&+ 
\begin{bmatrix}
Q_{l-1} + E_{l-1} Q_{l-1}^d E_{l-1}^T & E_{l-1}Q^d_{l-1} & 0\\
Q^d_{l-1}E_{l-1}^T & Q_{l-1}^d & 0\\
0 & 0 & Q_{l-1}^f
\end{bmatrix} \\
\label{e:P i|i-1 2}
& =
\begin{bmatrix}
P_{l|l-1}^x & P_{l|l-1}^{xd} & 0\\
P_{l|l-1}^{dx} & P_{l|l-1}^d & 0\\
0 & 0 & P_{l|l-1}^f
\end{bmatrix}
\end{align}
where 
\begin{align}
P_{l|l-1}^x & := A_{l-1} P_{l-1|l-1}^x A_{l-1}^T + E_{l-1}P_{l-1|l-1}^d E_{l-1}^T \nonumber \\
 &+ A_{l-1}P_{l-1|l-1}^{xd}E_{l-1}^T + E_{l-1}P_{l-1|l-1}^{dx}A_{l-1}^T \nonumber \\
 &+  Q_{l-1} + E_{l-1} Q_{l-1}^d E_{l-1}^T \nonumber \\
P_{l|l-1}^d & := P_{l-1|l-1}^d + Q_{l-1}^d \nonumber \\
P_{l|l-1}^{xd} &:= A_{l-1}P_{l-1|l-1}^{xd} + E_{l-1}P_{l-1|l-1}^d + E_{l-1}Q^d_{l-1} \nonumber \\
P_{l|l-1}^{dx} &:= P_{l-1|l-1}^{dx} A_{l-1}^T + P_{l-1|l-1}^d E_{l-1}^T + Q^d_{l-1}E_{l-1}^T \nonumber \\
P_{l|l-1}^f & := P_{0}^f + Q_{l-1}^f \nonumber  
\end{align}
Define
\begin{align}
\bar{C}_l := \bar{H}_{l} \bar{P}_{l|l-1} \bar{H}_{l}^T + R_l .
\end{align}
Substituting Eqs.~\eqref{e:wb fault} and (\ref{e:P i|i-1 2}) into the above equation, it follows that
\begin{align}
\bar{C}_l & = H_l P_{l|l-1}^x H_{l}^T + F_l P_{l|l-1}^f F_l^T + R_l
\end{align}
Consequently, the Kalman gain of the augmented filter can be calculated and partitioned as follows:
\begin{align}
\bar{K}_l &= \bar{P}_{l|l-1} \bar{H}_{l}^T \bar{C}_l^{-1} \nonumber \\
& = \begin{bmatrix}
P_{l|l-1}^x H_l^T\\
P_{l|l-1}^{dx} H_l^T \\
P_{l|l-1}^f F_l^T
\end{bmatrix}
\bar{C}_l^{-1}
\end{align}
If $P_0^f$ is chosen sufficiently large, then $P_{l|l-1}^f \approx P_{0}^f $ and $\bar{C}_l \approx F_l P_{0}^f F_l^T$. It follows that 
\begin{align}
\bar{K}_l = \begin{bmatrix}
P_{l|l-1}^x H_l^T \bar{C}_l^{-1} \\
P_{l|l-1}^{dx} H_l^T \bar{C}_l^{-1} \\
F_l^{-1}
\end{bmatrix}
\end{align}
Therefore, $K_l^f = F_l^{-1}$. It follows from Lemma~\ref{l:1} that $f_l$ can be estimated.     \rlap {$\qquad \Box$}
\end{pf}

\subsection{Unknown input estimation during $ l< k \leq l_e$}
\begin{thm}
Provided that $f_k$ has been estimated at $k=l$, $d_k$ can be estimated by the augmented fault filter of the extended \ac{DMAE} approach. 
\end{thm}
\begin{pf}
During this period, the augmented fault model matches the true fault scenario. Therefore, $i_{max,k}=2$, which means that the fault-free filter is reinitialized by the fault filter during this period.
Since this paper considers bias fault, $f_k$ is constant for $l< k \leq l_e$. 
Therefore, during this period, we can set:
\begin{align}
\hat{\bar{x}}_{k|k-1} &= \begin{bmatrix}
\hat{x}^*_{k|k-1} \\ \hat{f}_{l|l}
\end{bmatrix},
\bar{P}_{k|k-1} = \begin{bmatrix}
P^{*}_{k|k-1} & 0 \\ 0 & P_{l|l}^f
\end{bmatrix} \nonumber \\
\hat{\bar{x}}_{k|k} &= \begin{bmatrix}
\hat{x}^{*}_{k|k} \\ \hat{f}_{l|l}
\end{bmatrix},
\bar{P}_{k|k} = \begin{bmatrix}
P^{*}_{k|k} & 0 \\ 0 & P_{l|l}^f
\end{bmatrix},
\bar{K}_k = \begin{bmatrix}
K_k^{*} \\ 0
\end{bmatrix}, 
\end{align}
where 
\begin{align}
\hat{x}^{*}_{k|k-1} &:= \begin{bmatrix}
\hat{x}_{k|k-1} \\ \hat{d}_{k|k-1}
\end{bmatrix},
P^{*}_{k|k-1} := \begin{bmatrix}
P^x_{k|k-1} & P^{xd}_{k|k-1} \\
P^{dx}_{k|k-1} & P^d_{k|k-1}
\end{bmatrix}, \nonumber \\
\hat{x}^{*}_{k|k} &:= \begin{bmatrix}
\hat{x}_{k|k} \\ \hat{d}_{k|k}
\end{bmatrix},
P^{*}_{k|k} := \begin{bmatrix}
P^{x}_{k|k} & P^{xd}_{k|k} \\
P^{dx}_{k|k} & P^d_{k|k}
\end{bmatrix},
K^{*}_k := \begin{bmatrix}
K^x_k \\ K^d_k
\end{bmatrix} 
\end{align}
are updated by the normal Kalman filtering procedure. It can be seen that during this period, the estimation of the fault and the covariance are:
\begin{align}
\hat{f}_{k|k} = \hat{f}_{l|l}, \ P^f_{k|k} = P^f_{l|l},\  l< k \leq l_e
\end{align}
It can be inferred that the model of the fault filter is equivalent to:
\begin{align}
\label{e:xdot af}
x_{k+1} &= A_k x_k + E_k d_k + w_k\\
\label{e:y af}
y_{k} &= H_k x_k + F_k \hat{f}_{l|l} + v_k
\end{align}
As can be seen, the only unknown input is $d_k$ since the fault filter treats $f_k$ as a known input during this period. Since a known input does not affect the design of a filter \cite{Hou1998}, the convergence condition of this fault filter is the same as that of the fault-free filter based on Eqs.~\eqref{e:xdot ff} and \eqref{e:y ff}.

Therefore, $d_k$ can be estimated using the augmented fault filter under the same condition as for the model described by Eqs.~\eqref{e:xdot ff} and \eqref{e:y ff}.
\rlap {$\qquad \Box$}
\end{pf}

\subsection{Error analysis}
In the previous sections, it is assumed that $\hat{x}_{l-1|l-1}$ and $\hat{d}_{l-1|l-1}$ are unbiased. We analyze the estimation error of ${f}_{l}$ when $\hat{x}_{l-1|l-1}$ and $\hat{d}_{l-1|l-1}$ are biased. 

Through Eq.~\eqref{e:Kif}, Eq.~\eqref{e:fi est} can be further rewritten into
\begin{align}
\label{e:fi est 2}
\hat{f}_{l|l} &= f_l + F_l^{-1}e_l 
\end{align}
Substitute Eq.~\eqref{e:ei} into Eq.~\eqref{e:fi est 2}, it follows
\begin{align}
\hat{f}_{l|l} &= f_l + F_l^{-1} (H_l A_{l-1}\tilde{x}_{l-1|l-1} + H_l E_{l-1} \tilde{d}_{l-1|l-1}  \nonumber \\
 &+  H_l w_{l-1} + H_l E_{l-1}w_{d,l-1} + v_l)
\end{align}

The estimation error of ${f}_{l}$ as a function of $\tilde{x}_{l-1|l-1}$ and $\tilde{d}_{l-1|l-1} $ can be obtained as follows:
\begin{align}
\tilde{f}_{l|l} & = f_l - \hat{f}_{l|l} \\
&= F_l^{-1} (H_l A_{l-1}\tilde{x}_{l-1|l-1} + H_l E_{l-1} \tilde{d}_{l-1|l-1}  \nonumber \\
 &+  H_l w_{l-1} + H_l E_{l-1}w_{d,l-1} + v_l)
\end{align}
If $\tilde{x}_{l-1|l-1}$ and $\tilde{d}_{l-1|l-1} $ are unbiased, the expectation of $\tilde{f}_{l|l}$ is zero, which means the fault estimate is unbiased. If $\tilde{x}_{l-1|l-1}$ and $\tilde{d}_{l-1|l-1} $ are biased, assume
\begin{align}
\underaccent{\bar}{a} I \leq A_{l-1} \leq \bar{a} I,& \quad \underaccent{\bar}{f} I \leq F_{l-1} \leq \bar{f} I,\\
\underaccent{\bar}{h} I \leq H_{l-1} \leq \bar{h} I,& \quad \underaccent{\bar}{e} I \leq E_{l-1} \leq \bar{e} I, \\
\underaccent{\bar}{e}_x I \leq \tilde{x}_{l-1|l-1} \leq \bar{e}_x I,& \quad \underaccent{\bar}{e}_d I \leq \tilde{d}_{l-1|l-1}  \leq \bar{e}_d I, \\
\underaccent{\bar}{w} I \leq w_{l-1} \leq \bar{w} I,& \quad \underaccent{\bar}{w}_d I  \leq w_{d,l-1} \leq \bar{w}_d I ,\\
\underaccent{\bar}{v} I \leq v_{l-1} \leq \bar{v} I.  &
\end{align}
Then it follows that the fault estimation error is bounded by the following:
\begin{align}
\big[\frac{  \underaccent{\bar}{h} ( \underaccent{\bar}{a}  \underaccent{\bar}{e}_x +   \underaccent{\bar}{e}  \underaccent{\bar}{e}_d + \underaccent{\bar}{w} + \underaccent{\bar}{e} \underaccent{\bar}{w}_d ) + \underaccent{\bar}{v} }{ \bar{f} }  ,  \frac{  \bar{h} ( \bar{a}  \bar{e}_x +   \bar{e}  \bar{e}_d + \bar{w} + \bar{e} \bar{w}_d ) + \bar{v} }{ \underaccent{\bar}{f} } \big]
\end{align}

\subsection{Discussion }
For the model given in Eqs.~\eqref{e:xdot ff} and \eqref{e:y ff}, the convergence condition for time-invariant case has been given by Darouach et al. \cite{Darouach1995}, which is given as follows:
\begin{align}
\text{rank} \begin{bmatrix}
zI-A & & -E\\ H& &  0
\end{bmatrix}
=n+ n_d,
\forall z\in \mathcal{C}, |z| \ge 1
\end{align}
This convergence condition is also required by traditional unknown input filters such as those in Darouach, Zasadzinski and Boutayeb \cite{Darouach2003} and Cheng et al. \cite{Cheng2009}.

The system considered in this paper is linear and the noise is assumed to be Gaussian. If the system is nonlinear, the \ac{DMAE} should be extended using Unscented Kalman Filters \cite{Julier1997,Lu2015} or particle filters \cite{Gordan1993,Doucet2000,Verma2004}. If the system noise is non-Gaussian, then it should be extended by making use of particle filters \cite{Gordan1993,Doucet2000,Verma2004}. However, this is out of the scope of the present paper.

\section{Illustrative examples with comparison to existing methods}
\label{s:5}
In this section, two examples similar to that in \cite{Park2000}, \cite{Darouach2003} and \cite{Hsieh2009} are provided to demonstrate the performance of the extended \ac{DMAE} approach. Note that both $E$ and $F$ are of full rank in this example.

\begin{figure}
\subfigure[True and estimated states, example 1]{
\includegraphics[width = 0.5\textwidth]{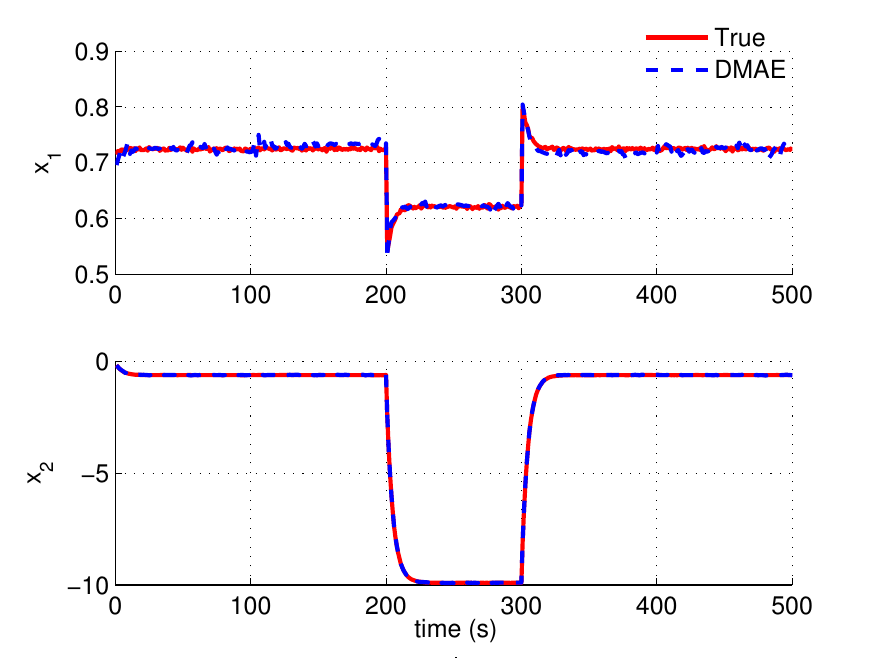}
\label{f:state df same kind}
}
\subfigure[True and estimated disturbances, example 1]{
\includegraphics[width = 0.5\textwidth]{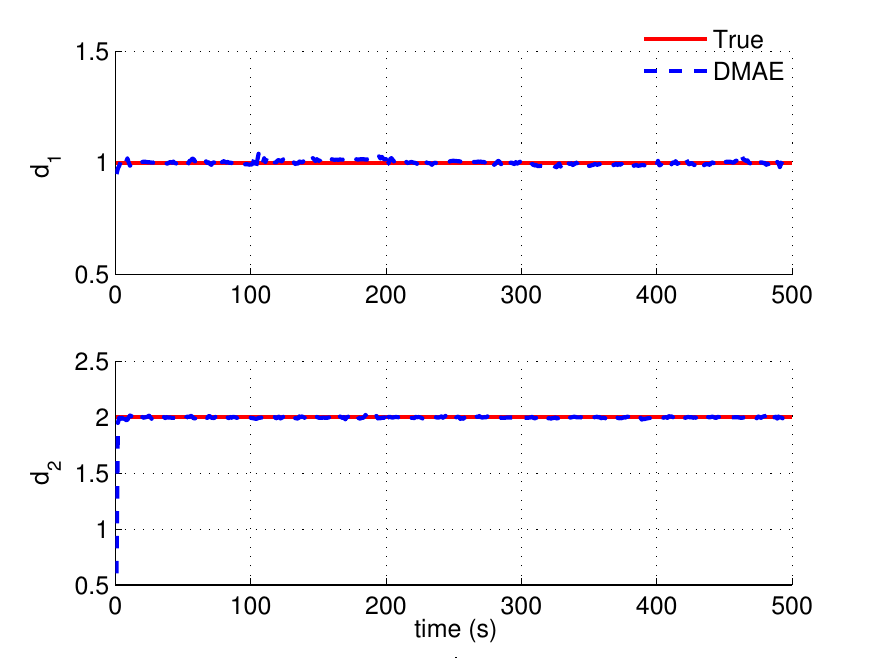}
\label{f:dist df same kind}
}
\subfigure[True and estimated faults, example 1]{
\includegraphics[width = 0.5\textwidth]{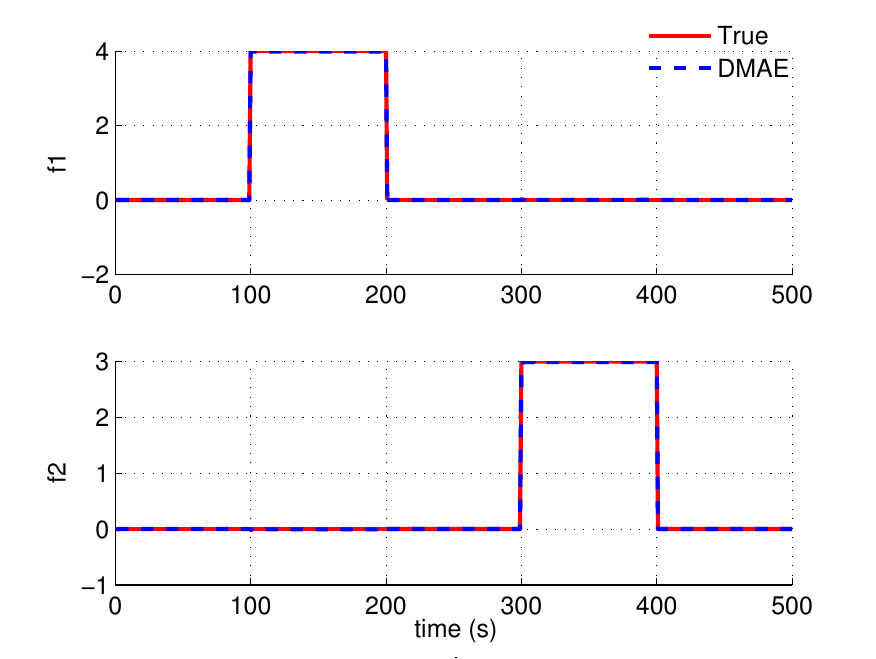}
\label{f:fault df same kind}
}
\caption{Results of the \ac{DMAE} approach, example 1}
\label{f:df same kind}
\end{figure}

The system is described by model (\ref{e:xdot}) and (\ref{e:y}) where 
\begin{align}
A &= \begin{bmatrix}
-0.0005 & -0.0084 \\
0.0517  & 0.8069
\end{bmatrix},
B = \begin{bmatrix}
0.1815\\
1.7902
\end{bmatrix}, \\
E &= \begin{bmatrix}
0.629 & 0\\
0   &  -0.52504
\end{bmatrix},
H =\begin{bmatrix}
1 & 0\\
0 & 1
\end{bmatrix},
F= \begin{bmatrix}
1 & 0\\
0 & 1
\end{bmatrix},\\
Q &= \begin{bmatrix}
0.002^2 & 0\\
0 & 0.002^2
\end{bmatrix},
R= \begin{bmatrix}
0.01^2 & 0\\
0 & 0.01^2
\end{bmatrix}
\end{align}
The input $u_k$ is: $u_k= -0.5$ when $200<k \le 300$, otherwise $u_k = 0.5$. $f_k$ is given by the red solid lines in Fig.~\ref{f:fault df same kind}. 
It can be noted that the number of unknown inputs in \cite{Park2000}, \cite{Darouach2003} and \cite{Hsieh2009} is $n_d$ ($n_d=2$) while this paper deals with $2n_d$ unknown inputs.

In both examples, since $E'_k = [E_k\ 0]$, $F'_k = [0\ F_k]$, condition~(\ref{e:existence condition}) is not satisfied. In addition, rank $y_k < $ rank $d_k'$. Consequently, all the unknown input decoupled filters in the introduction are not applicable to solve the problem, except for special cases when $d_k=0$ or $f_k =0$. $N$ in Eq.~\eqref{e:innovation covariance} is set to be 10. In both examples, $Q_k^f=0$, $Q_k^d$ is updated by the main diagonal of the matrix given in~\eqref{e:Qd compensation}, $x_0^f = [10^{-3}, 10^{-3}]^T, P_0^f = 10^{2}I$.

\textbf{Example 1}. In this example, $d_k$ is a constant bias vector, which is shown by the red solid lines in Fig.~\ref{f:dist df same kind}. The condition (\ref{e:existence condition}) is not satisfied. Therefore, traditional unknown input filters, which require the satisfaction of condition (\ref{e:existence condition}), can not be implemented.

The extended \ac{DMAE} approach is implemented. 
The true and estimated $p_{n\!f}$ and $p_{a\!f}$ using the extended \ac{DMAE} approach are well matched.
The probability-weighted estimates of $x_k$, $d_k$, which are calculated using Eq.~\eqref{e:state pdf}, are shown in Fig.~\ref{f:state df same kind} and \ref{f:dist df same kind}, respectively. The probability-weighted estimate of $f_k$ (calculated using Eq.~\eqref{e:fault pdf}) is shown in Fig.~\ref{f:fault df same kind}. As can be seen, $x_k$, $d_k$ and $f_k$ can all be estimated.

\begin{figure}
\subfigure[True and estimated model probabilities, case 2, example 2]{
\includegraphics[width = 0.5\textwidth]{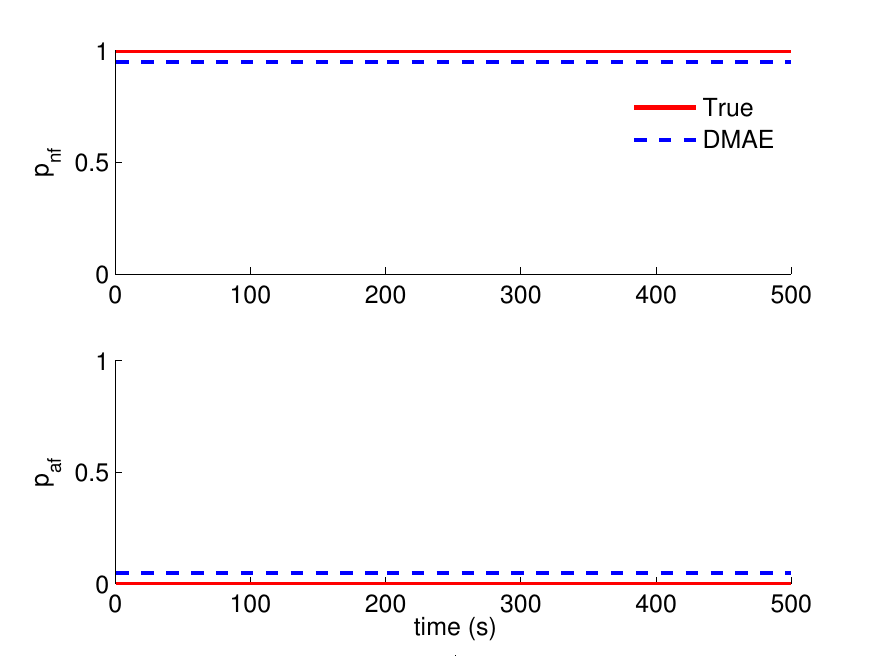}
\label{f:pdf d}
}
\subfigure[True and estimated disturbances, case 2, example 2]{
\includegraphics[width = 0.5\textwidth]{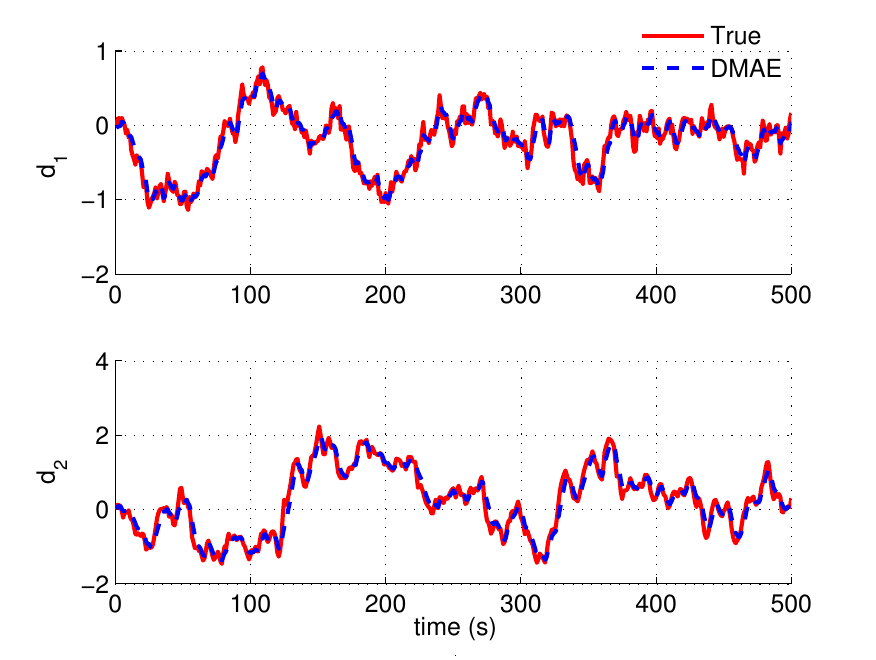}
\label{f:dist d}
}
\caption{Results of the \ac{DMAE} approach, case 2, example 2}
\label{f:d}
\end{figure}

\textbf{Example 2}. In this example\footnote{The implementation of this work is available at: \\ \url{https://www.researchgate.net/profile/Peng_Lu15/publications?pubType=dataset} }, the disturbances, which are taken from \cite{Mclean1990}, are stochastic.
$d_k = \begin{bmatrix}
d_{1,k}\\ d_{2,k}
\end{bmatrix}$ is generated using the following model \cite{Mclean1990}:
\begin{align}
\begin{bmatrix}
{d_{i,k}}\\
{d_{i,k}^{'}}
\end{bmatrix}
& =
\begin{bmatrix}
0 & 1\\
-\frac{V^2}{L_{gi}^{2}} & -2\frac{V}{L_{gi}}
\end{bmatrix}
\begin{bmatrix}
{d_{i,k-1}}\\
{d_{i,k-1}^{'}}
\end{bmatrix} \nonumber  \\
 &+ \begin{bmatrix}
\sigma_i \sqrt{\frac{3V}{L_{gi}}} \\
(1-2\sqrt{3}) \sigma_i \sqrt{ (\frac{V}{L_{gi}})^3 }
\end{bmatrix}
w'_{d,k}, \ i=1,2
\end{align}
where $V=35$, $\sigma_1 = 0.5$, $\sigma_2=0.8$, $L_{g1} = 2500$, $L_{g2} = 1500$ and $w'_{d,k} \sim N(0,1)$. The generated $d_k$ is shown by the red solid lines in Fig.~\ref{f:dist d}. It should be noted that the \ac{DMAE} approach still models $d_k$ as a random walk process since $d_k$ is treated as an unknown input.

Three cases are considered for this example. The first two cases are special cases. In these two cases, the existence condition (\ref{e:existence condition}) is satisfied. Therefore, some of the approaches mentioned in the introduction can still be used.
\begin{case}
$d_k=0$, $f_k \neq 0$
\end{case}

\begin{figure}
\includegraphics[width = 0.5\textwidth]{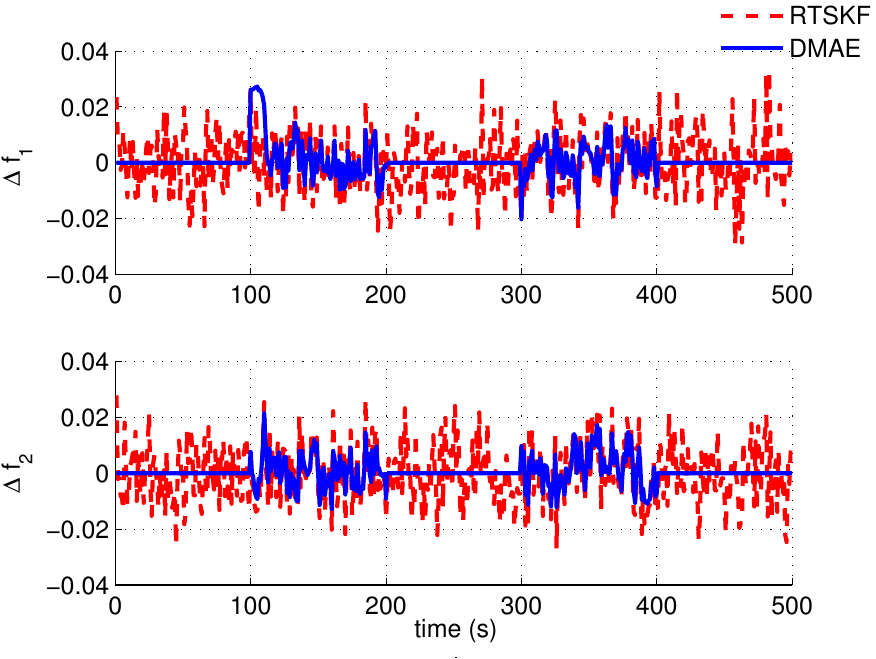}
\caption{Errors of estimation of $f_1$ and $f_2$ using the \ac{RTSKF} and the \ac{DMAE} approach, case 1, example 2}
\label{f:fault f R3SKF}
\end{figure}

In this case, $E_k$ is a zero matrix. Therefore, condition (\ref{e:existence condition}) is satisfied. 
The probability-weighted estimate of $f_k$ using the extended \ac{DMAE} is the same as in Fig.~\ref{f:fault df same kind}.
The \ac{RTSKF} in Gillijns and De Moor \cite{Gillijns2007} is also applied and the errors of estimation of $f_k$ compared to the \ac{DMAE} are shown in Fig.~\ref{f:fault f R3SKF}. In addition, particle filters \cite{Gordan1993,Doucet2000} are also applied. The model used for estimation of $f_k$ is also the random walk. 100 particles are used.
The \acp{RMSE} of estimation of $f_1$ and $f_2$ using the \ac{RTSKF}, the particle filter \cite{Gordan1993,Doucet2000} and the extended \ac{DMAE} are shown in Table~\ref{t:RMSE}.

\begin{table}
\caption{RMSEs of the fault and disturbance estimation for Example 2}
\centering
\begin{tabular}{ cccccc }
\hline \hline
   & Methods & $d_1$ & $d_2$ & $f_1$ & $f_2$\\ \hline
\multirow{2}{*}{Case 1}
 & RTSKF \cite{Gillijns2007} & -  & -   & 0.0103   &  0.0102   \\
 & PF \cite{Gordan1993,Doucet2000} & -   &  -   & 0.1549  &  0.1496   \\ 
 & DMAE & -   &  -   & 0.0060  &  0.0047   \\ \hline
\multirow{2}{*}{Case 2}
 & OTSKF \cite{Heish1999} &  0.0697  &  0.1442  & -  &  -  \\
 & PF \cite{Gordan1993,Doucet2000} &  0.1088  &  0.2035    & -  &  -  \\ 
 & DMAE &  0.0709  &  0.1459   & -  &  -  \\ \hline
\multirow{2}{*}{Case 3}
 & {\color{black}\cite{Gillijns2007,Gillijns2007c,Heish1999,Gordan1993,Doucet2000,Hsieh2000}} & N/A   &  N/A  & N/A  & N/A   \\
 & DMAE & 0.0845   & 0.1655    &  0.0230  & 0.0283  \\ \hline
\end{tabular}
\label{t:RMSE}
\end{table}

\begin{case}
$d_k\neq 0$, $f_k = 0$
\end{case}
In this case, $F_k$ is a zero matrix. Therefore, condition (\ref{e:existence condition}) is also satisfied. 
The true and estimated $p_{n\!f}$ and $p_{a\!f}$ using the extended \ac{DMAE} approach are shown in Fig.~\ref{f:pdf d}.
The probability-weighted estimate of $d_k$ is presented in Fig.~\ref{f:dist d}. The results using the methods in Heish \cite{Hsieh2000}, Heish and Chen \cite{Heish1999}, and Gillijns and De Moor \cite{Gillijns2007c}, are similar to that of the \ac{DMAE}. Particle filter is also applied. The model used for estimation of $d_k$ is the random walk.
The \acp{RMSE} of estimation of $d_1$ and $d_2$ using the \ac{OTSKF} in Heish \cite{Heish1999}, the particle filter \cite{Gordan1993,Doucet2000} and the extended \ac{DMAE} are shown in Table~\ref{t:RMSE}.

\begin{case}
$d_k \neq 0$, $f_k \neq 0$
\end{case}
In this case, condition (\ref{e:existence condition}) is not satisfied. Thus, all the conventional filters mentioned in the introduction are not applicable.

\begin{figure}
\includegraphics[width = 0.5\textwidth]{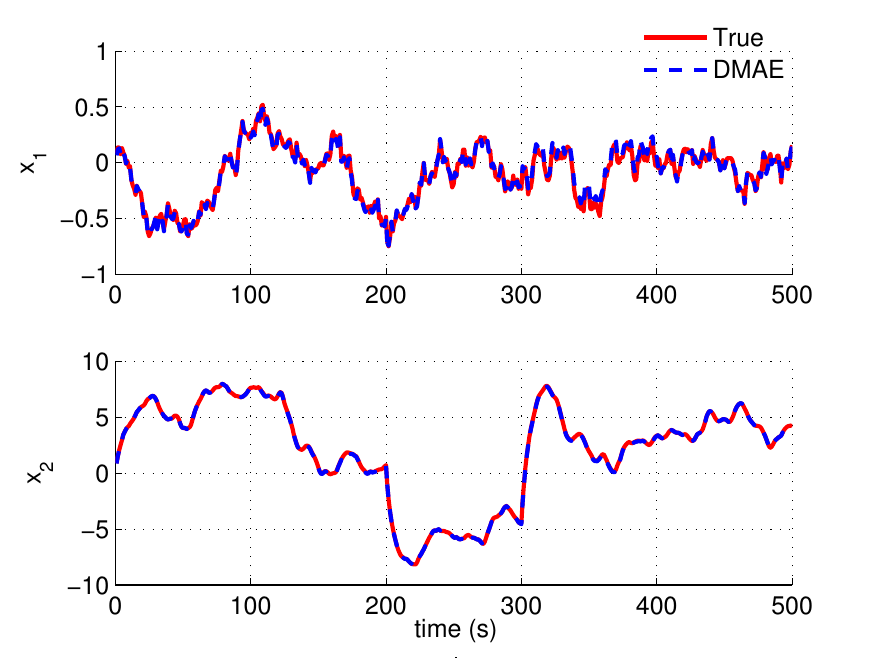}
\caption{True and estimated states, case 3, example 2}
\label{f:state df}
\end{figure}

The true and estimated $p_{n\!f}$ and $p_{a\!f}$ using the extended \ac{DMAE} approach are also well matched.
The probability-weighted estimates of $x_k$, is shown in Fig.~\ref{f:state df}. The probability-weighted estimates of $d_k$ and $f_k$ are the same as in Figs.~\ref{f:dist d} and \ref{f:fault df same kind} respectively.
It can be seen that despite the fact that the existence condition for traditional unknown-input decoupled filters is not satisfied, $x_k$, $d_k$ and $f_k$ can all be estimated using the extended \ac{DMAE} approach. The \acp{RMSE} of the estimation of $d_k$ and $f_k$ using the extended \ac{DMAE} approach are shown in Table~\ref{t:RMSE}.

\begin{figure}
\subfigure[Sensitivity with respect to $Q_k$ errors, case 3, example 2]{
\includegraphics[width = 0.5\textwidth]{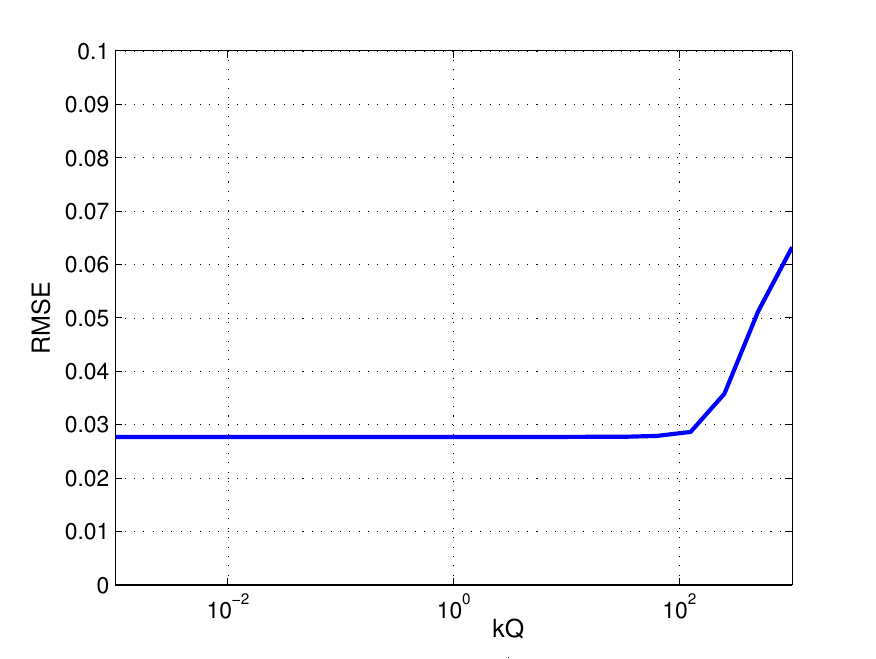}{}
\label{f:sensitivity Q}
}
\subfigure[Sensitivity with respect to $R_k$ errors, case 3, example 2]{
\includegraphics[width = 0.5\textwidth]{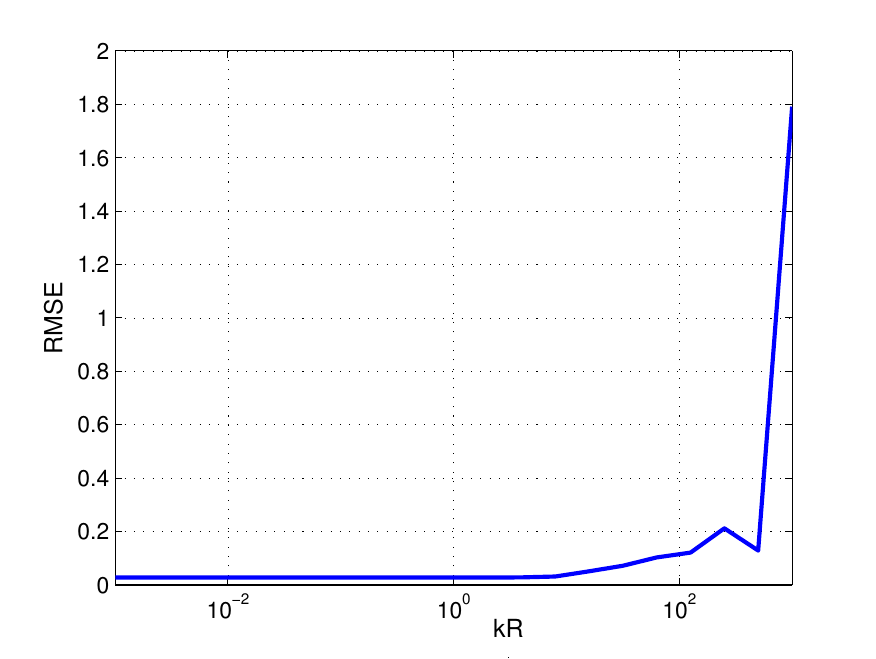}{}
\label{f:sensitivity R}
}
\caption{Sensitivity of the fault estimation using the \ac{DMAE} approach with respect to $Q_k$ and $R_k$ errors, case 3, example 2}
\label{f:sensitivity}
\end{figure}

Finally, the sensitivity of the \ac{DMAE} with respect to errors in $Q_k$ and $R_k$ is discussed. To demonstrate the sensitivity with respect to errors in $Q_k$, $R_k$ is fixed and $Q_k$ is multiplied with a coefficient $k_Q$. The sensitivity result of the \ac{RMSE} of fault estimation with $k_Q$ ranging from $10^{-3}$ to $10^3$ is shown in Fig.~\ref{f:sensitivity Q}.
To show the sensitivity with respect to $R_k$ errors, $Q_k$ is fixed and $R_k$ is multiplied with a coefficient $k_R$. The sensitivity result of the \ac{RMSE} of fault estimation with $k_R$ ranging from $10^{-3}$ to $10^3$ is shown in Fig.~\ref{f:sensitivity R}.

It can be seen from Fig.~\ref{f:sensitivity Q} and \ref{f:sensitivity R} that the minimum \acp{RMSE} are obtained when $k_Q=1$ or $k_R=1$. However, it is also noted that the extended \ac{DMAE} approach is more sensitive to $R_k$ errors. The \ac{RMSE} of the fault estimation increases to 0.063 when $Q_k$ is multiplied with $10^3$ and increases to $1.79$ when $R_k$ is multiplied with $10^3$. This is expected since in section~\ref{s:3}, the process noise $\bar{w}_{n\!f,k}$ is adapted while the output noise $v_{k}$ is not adapted. Therefore, selection of $R_k$ should be performed with more caution.

\section{Conclusion}
\label{s:6}
In this paper, the unknown input decoupling problem is extended to the case when the existence condition of traditional unknown input filters is not satisfied. It is proved that the states, disturbances and faults can be estimated using an extended \ac{DMAE} approach which does not require the existence condition. Therefore, it can be applied to a wider class of systems and applications. Two illustrative examples demonstrate the effectiveness of the extended \ac{DMAE} approach. Future work would consider extending the \ac{DMAE} to deal with systems with non-Gaussian noise.

\bibliographystyle{plain}        
\bibliography{lite_UI}    

\end{document}